\documentclass{aa}    
\usepackage{graphicx,url,twoopt,natbib}
\usepackage[varg]{txfonts}           
\usepackage{hyperref}                
\hypersetup{
  colorlinks=false,   
  urlcolor=blue,     
  linkcolor=red,     
}

\bibpunct{(}{)}{;}{a}{}{,}    

\makeatletter
\renewcommand*\aa@pageof{, page \thepage{} of \pageref*{LastPage}}
\makeatother

\begin{document}  

\titlerunning{Dynamical mass constraints with {\it Gaia} DR2}
\authorrunning{Calissendorff \& Janson}

\title{Improving dynamical mass constraints for intermediate-period substellar companions using {\it Gaia} DR2}

\author{Per Calissendorff$^{1}$\and Markus Janson$^1$
                                }

\institute{
        Department of Astronomy, Stockholm University, Stockholm, Sweden\\
        e-mail: {\bf per.calissendorff@astro.su.se}
}        

\date{Received 26 April 2018 / Accepted 15 June 2018}

\abstract{
The relationship between luminosity and mass is of fundamental importance for direct imaging studies of brown dwarf and planetary companions to stars. In principle this can be inferred from theoretical mass-luminosity models; however, these relations have not yet been thoroughly calibrated, since there is a lack of substellar companions for which both the brightness and mass have been directly measured. One notable exception is GJ 758 B, a brown dwarf companion in a $\sim$20~AU orbit around a nearby Sun-like star, which has been both directly imaged and dynamically detected through a radial velocity trend in the primary. This has enabled a mass constraint for GJ 758 B of 42$^{+19}_{-7}$~$M_{\rm Jup}$. Here, we note that {\it Gaia} is ideally suited for further constraining the mass of intermediate-separation companions such as GJ 758 B. A study of the differential proper motion, $\Delta \mu$, with regards to HIPPARCOS is particularly useful in this context, as it provides a long time baseline for orbital curvature to occur. By exploiting already determined orbital parameters, we show that the dynamical mass can be further constrained to $42.4^{+5.6}_{-5.0}\, M_{\rm Jup}$ through the {\it Gaia}-HIPPARCOS $\Delta \mu$ motion. We compare the new dynamical mass estimate with substellar evolutionary models and confirm previous indications that there is significant tension between the isochronal ages of the star and companion, with a preferred stellar age of $\leq 5$ Gyr while the companion is only consistent with very old ages of $\geq 8$ Gyr.
}  

\keywords{astrometry --- brown dwarfs --- stars: kinematics and dynamics}
\maketitle





\section{Introduction} \label{sec:intro}
\noindent Multiple systems of stars and substellar companions prove to be excellent benchmarks to be compared with evolutionary models, as their dynamical motions can yield model-independent masses provided that their orbits can be constrained. The systems become even more valuable if their ages can be accurately determined, as they provide a means to calibrate the evolutionary models directly from what is observed. For low-mass multiple systems, evolutionary models generally underpredict the mass by a factor of $10 - 30\%$ compared to the dynamical estimates \citep[e.g.][]{Montet+15, Calissendorff}. In the even-lower-mass regimes, towards substellar brown dwarfs,  detected multiple systems with constrained orbits become less frequent, leading to a higher degree of uncertainty in the predictions by the models \citep[e.g.][]{Dupuy+09}.

Astrometric monitoring of substellar companions on long orbits typically requires a good portion of the orbit to be covered in order to yield any robust results. Nevertheless, in some cases, the companion imposes a strong gravitational pull on its host,  sufficient to produce a measurable Doppler shift, which can aid in the determination of the orbit and mass of the system. Although the companion may not always be resolvable, precise astrometric measurements like those of HIPPARCOS \citep{Perryman+97} and {\it Gaia} \citep{Prusti+16} may reveal insights to the small astrometric perturbation the companions have on their host star. Given the timescale between two proper motion measurements, and the acceleration of the companion, a shift in proper motion for the entire system can potentially be detected and used to infer more stringent dynamical mass solutions for suitable targets.

GJ 758 is a bright solar-type $\approx$ G8 star with $V = 6.3$ mag, located at a distance of $d = 15.76 \pm 0.9$ pc \citep{van Leeuwen}, and slightly enhanced metallicity of $[{\rm Fe}/{\rm H}] = 0.18 \pm 0.05$ \citet{Vigan+16}. While the age of the system has long been uncertain, with estimations ranging from $\approx 40$ Myr to $\approx 10$ Gyr, most estimations point towards older ages of several Gyrs \citep[e.g.][]{Janson+11}. Combining HIPPARCOS parallax and proper motions by \citet{van Leeuwen} with absolute radial velocity by \citet{Nidever+02} yields space velocities of $(U, V, W) = (-21.1 \pm 0.2, -14.1 \pm 0.5, -3.0 \pm 0.2)$ km$^{-{\rm s}}$, which places the system close to the $\sim$ 40 Myr-old Argus association \citep[BANYAN II tool,][]{Gagne+14}. When taking into account information from both photometry and kinematics, the updated BANYAN $\Sigma$-tool instead estimates with a $99.4\%$ probability that GJ 758 is an old field star \citep{Gagne+18}. However, it should be pointed out that the Argus association is not included in the BANYAN $\Sigma$ analysis, as the association itself may be contaminated and not be considered as coeval \citep{Bell+15}. Chemical tagging of the star further supports an older age for the system and rules out Argus association membership. For young stars, lithium can be used as a sensitive age indicator, typically traced by the Li 6708\,\AA~resonance line. The lack of lithium observed by \citet{Vigan+16} sets a stringent lower limit of the age of $\sim$ 600 Myr; at this age, detectable amounts of lithium should have vanished for stars with the colours exhibited by GJ 758. Future asteroseismology studies may help to refine the age estimate further.

The companion, GJ 758B, was first detected with the Subaru/HICIAO by \citet{Thalmann+09}, and is likely to be a late T-type brown dwarf companion. Depending on the adopted age of the system, evolutionary models predict masses for the companion of between 10 and 40 $M_{\rm Jup}$ \citep[][]{Janson+11, Nilsson+17, Bowler+18}. However, evolutionary models are highly uncertain at these low masses. The companion is close enough to its host star that its gravitational influence is measurable with radial velocity (RV), yet well enough separated that both components can be resolved and characterised with direct imaging. The imaging combined with the long-term RV trends observed make the companion an especially useful benchmark object, as only a handful of similar systems have been detected so far.

By combining high-contrast imaging astrometric data of GJ 758B and the observed radial velocity trends, \citet{Bowler+18} employ a Monte Carlo Markov Chain (MCMC) simulation in order to estimate a dynamical mass for the companion of $42^{+19}_{-7} M_{\rm Jup}$, independent of the age of the system. Assuming that both GJ 758 and its companion are coeval, isochrones from evolutionary models then predict from the mass of the companion that the system is likely to be older than a few gigayears.

\section{Method} \label{sec:method}
\noindent Using the marginalised posteriors of the fitted orbital parameters from the MCMC analysis made by \citet{Bowler+18}, we construct our own simulation in order to predict which orbits are compatible with the proper motion of the system measured by {\it Gaia} in DR2 \citep{Brown+18}. We adopt the distance parallax $\pi_{\rm HIPPARCOS} = 63.45 \pm 0.35$ mas from \citet{van Leeuwen} in order to be consistent with the simulations by \citet{Bowler+18}. For the proper motions of GJ 758 we adopt the HIPPARCOS values from \citet{Perryman+97} of $\mu_{\rm RA, HIPPARCOS} = 82.04 \pm 0.54$ mas yr$^{-1}$ and $\mu_{\rm DEC, HIPPARCOS} = 162.92 \pm 0.52$ mas yr$^{-1}$, which are compared to the \emph{Gaia} proper motions, $\mu_{\rm RA, \emph{Gaia}} = 81.80 \pm 0.03$ mas yr$^{-1}$ and $\mu_{\rm DEC, \emph{Gaia}} = 160.39 \pm 0.04$ mas yr$^{-1}$.

First, we extract the posterior distributions from the histograms in \citet{Bowler+18} using the online webtool {\texttt WebPlotDigitizer}\footnote{\url{https://automeris.io/WebPlotDigitizer}} \citep{Rohatgi}. For each parameter, we then calculate the cumulative density function (CDF) and take $N = 20\, 000$ uniformly distributed random points between 0 and 1 that we translate using the CDFs into trial points for each parameter. We calculate the orbit for each set of orbital parameters simulated in this way and calculate the positional angle (PA) and relative separation at the epochs of the astrometric data points. We then compare the predicted values to the relative astrometry of \citet{Bowler+18} with a $\chi^2$-test weighted by the errors in the astrometry and save the $n_{\rm s} = 100$ best orbits. This procedure is required in order to remove unfeasible combinations of orbital parameters related to covariances in the data. The procedure is repeated $n_{\rm t} = 200$ times until we have 20 000 saved orbits.

From the saved orbits, we can predict the position of the companion during the HIPPARCOS and {\it Gaia} epochs. Using the HIPPARCOS Intermediate Astrometric Data tool\footnote{\url{https://www.cosmos.esa.int/web/hipparcos/java-tools/intermediate-data}}, we identify the orbits (and thereby epochs) during which  GJ 758 was observed. This gives us 21 unique orbits in which the target was observed. We then estimate the position of the companion for each observed epoch and calculate the proper motion value for the companion as the difference in position divided by the time between the observations. We thereafter take the median value for the proper motion of the companion at the HIPPARCOS epoch of 1991.25 as $\mu_{\rm c, HIPPARCOS}$. Using instead the first and last observed epochs divided by the time difference yields similar proper motions for the simulated orbits, with a median discrepancy of $\lesssim 20\,\mu$mas. For the {\it Gaia} epochs, we use the {\it Gaia} Observation Forecast Tool \footnote{\url{https://www.cosmos.esa.int/web/gaia/tools}} to estimate the orbits and dates during which GJ 758 was observed. We find 18 unique predicted epochs within the time-frame of 25 July 2014 (10:30 UTC) and 23 May 2016 (11:35 UTC) on which the {\it Gaia} DR2 data is based, and note that this is 2 more than the 16 epochs that were used for calculating the astrometric solution during the visible periods. Without knowing which exact epochs were used in the astrometry, we include all of them in our calculations and employ the same methods as before to estimate the companion proper motion at the {\it Gaia} epoch at 2015.50 as $\mu_{\rm c, {\it Gaia}}$. We then calculate the difference in proper motion between the two epochs for the companion as

\begin{equation}\label{eq:dmuc}
\Delta \mu_{\rm c} = \mu_{\rm c, \emph{Gaia}} - \mu_{\rm c, HIPPARCOS},
\end{equation}

and scale the difference in proper motion for the companion up to the star as
\begin{equation}\label{eq:dmus}
\Delta \mu_* =  -\frac{\Delta \mu_{\rm c}}{\frac{M_\odot}{M_{\rm c}}+1},
\end{equation}

where the $+1$ is added to the denominator to account for the fact that in the companion imaging we measure the companion position relative to the photocentre instead of relative to the centre of mass. The contrast in brightness between the primary and the companion causes the astrometric position of the system to be unaffected by the presence of the companion. As the companion is expected to be very dark in the {\it Gaia} band, the photocentre is centred on the primary.

\begin{figure}
\includegraphics[width=\columnwidth]{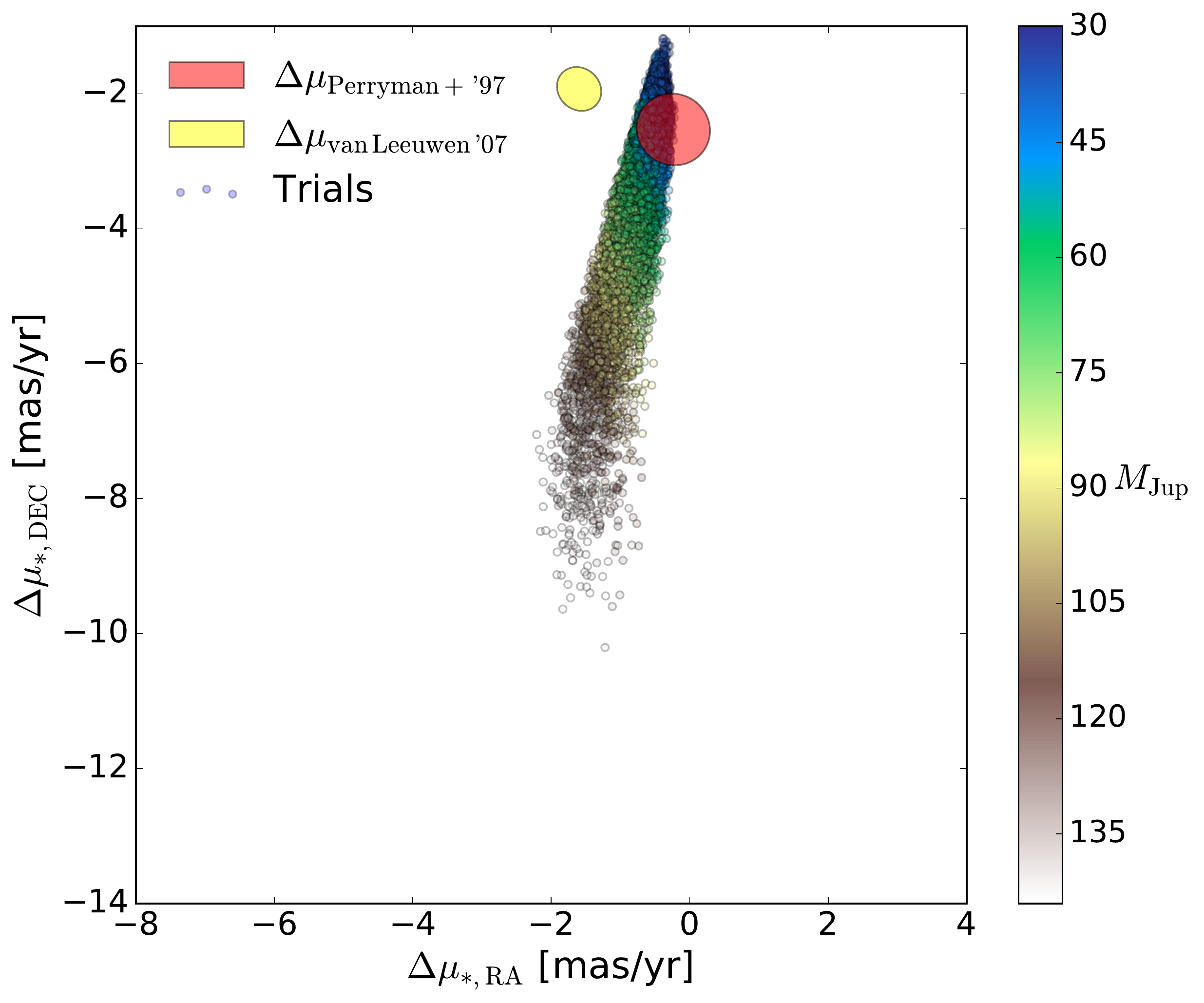}
\caption{The difference in proper motions between the HIPPARCOS and the {\it Gaia} epochs. Each dot represents a trial from our $N=20\, 000$ saved simulation trial points, and the red and yellow encircled areas show which values are compatible within $1-\sigma$ of the actual measured value when applying the HIPPARCOS proper motions from \citet{Perryman+97} and \citet{van Leeuwen}, respectively (see Sect.~\ref{sec:method} for discussion). Higher-mass companions induce stronger $\Delta \mu$, as indicated by the colour scheme. The colours of the encircled areas of the measured proper motions are not reflected by the defined colour scheme in mass, and serve only to differentiate between the two proper-motion solutions.}
\label{fig:pm}
\end{figure}

\begin{figure}
\includegraphics[width=\columnwidth]{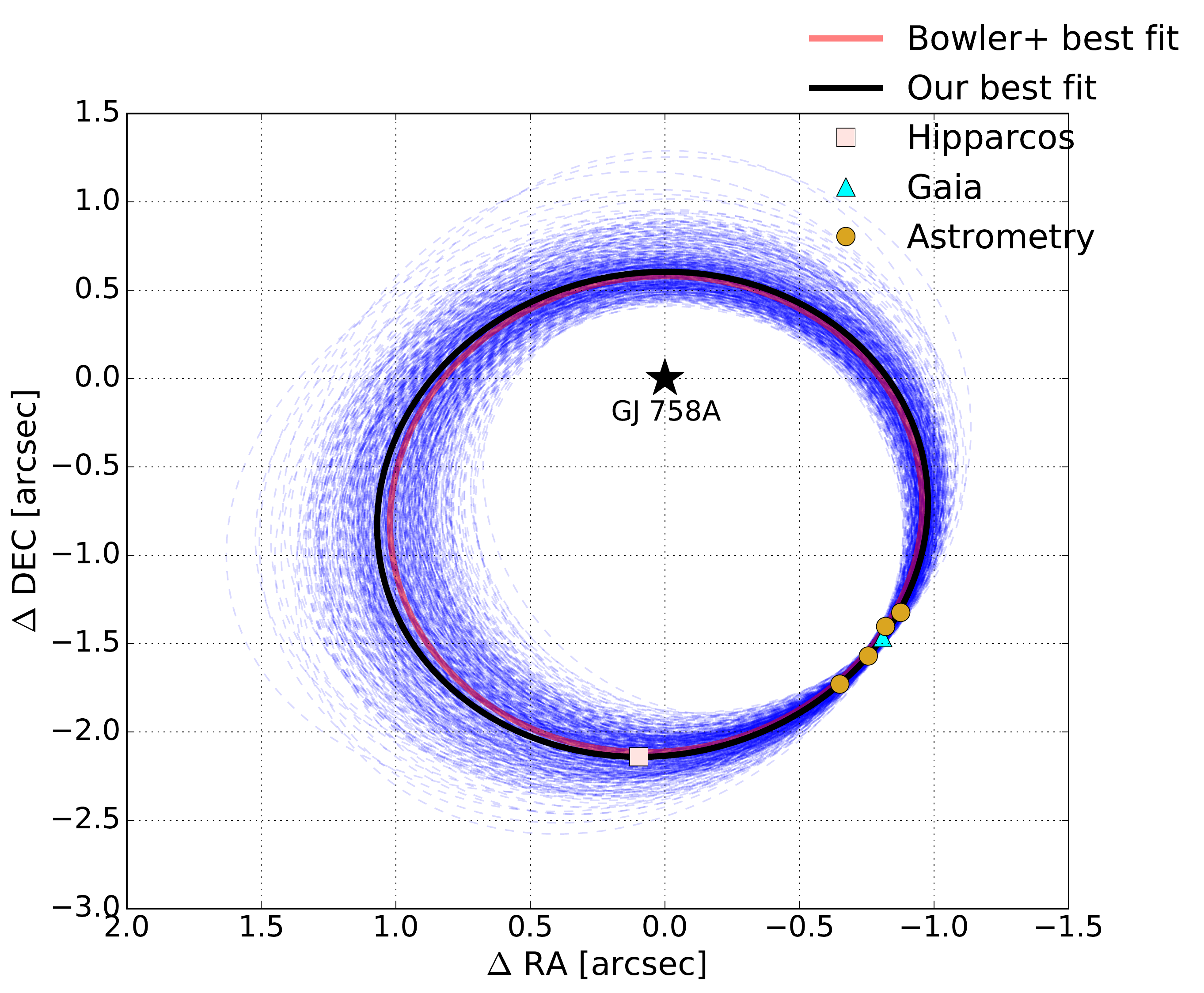}
\caption{Fitted orbits of the predicted values that correspond to the measured $\Delta \mu$, given by the blue dashed lines in the plot. The red line is the best-fit orbit from \citet{Bowler+18} and the black solid line shows the orbit for the best-fit orbit from the allowed $\Delta\mu$ values. The yellow circles show the relative astrometric measurement from the literature, ranging epochs from 2010.5 to 2017.8. The epochs at 1991.25 and 2015.50 depicted by the rose-tinted square and the cyan triangle represent the HIPPARCOS and {\it Gaia} epochs, respectively.}
\label{fig:orb}
\end{figure}

This provides us with a set of 20 000 predicted values for the difference in proper motion between the two epochs, as shown in Fig.~\ref{fig:pm}. The median values for the predicted difference in proper motions are $\Delta\mu_{*, {\rm RA}} = -0.61$ mas yr$^{-1}$ and $\Delta\mu_{*, {\rm DEC}} = -2.80$ mas yr$^{-1}$ , respectively. We then compare these predicted values with the actual difference using the new {\it Gaia} DR2 value. The predicted $\Delta \mu_*$ values that coincide within the uncertainty of measured values of $\Delta \mu_{\rm RA} = -0.24 \pm 0.54$ mas yr$^{-1}$ and $\Delta \mu_{\rm DEC} = -2.53 \pm 0.52$ mas yr$^{-1}$ provide us with a family of simulated trial points that correspond to allowed orbits, which are plotted in Fig.~\ref{fig:orb}. Approximately one third of the 20 000 simulated trial values were accepted.

\begin{figure}
\includegraphics[width=\columnwidth]{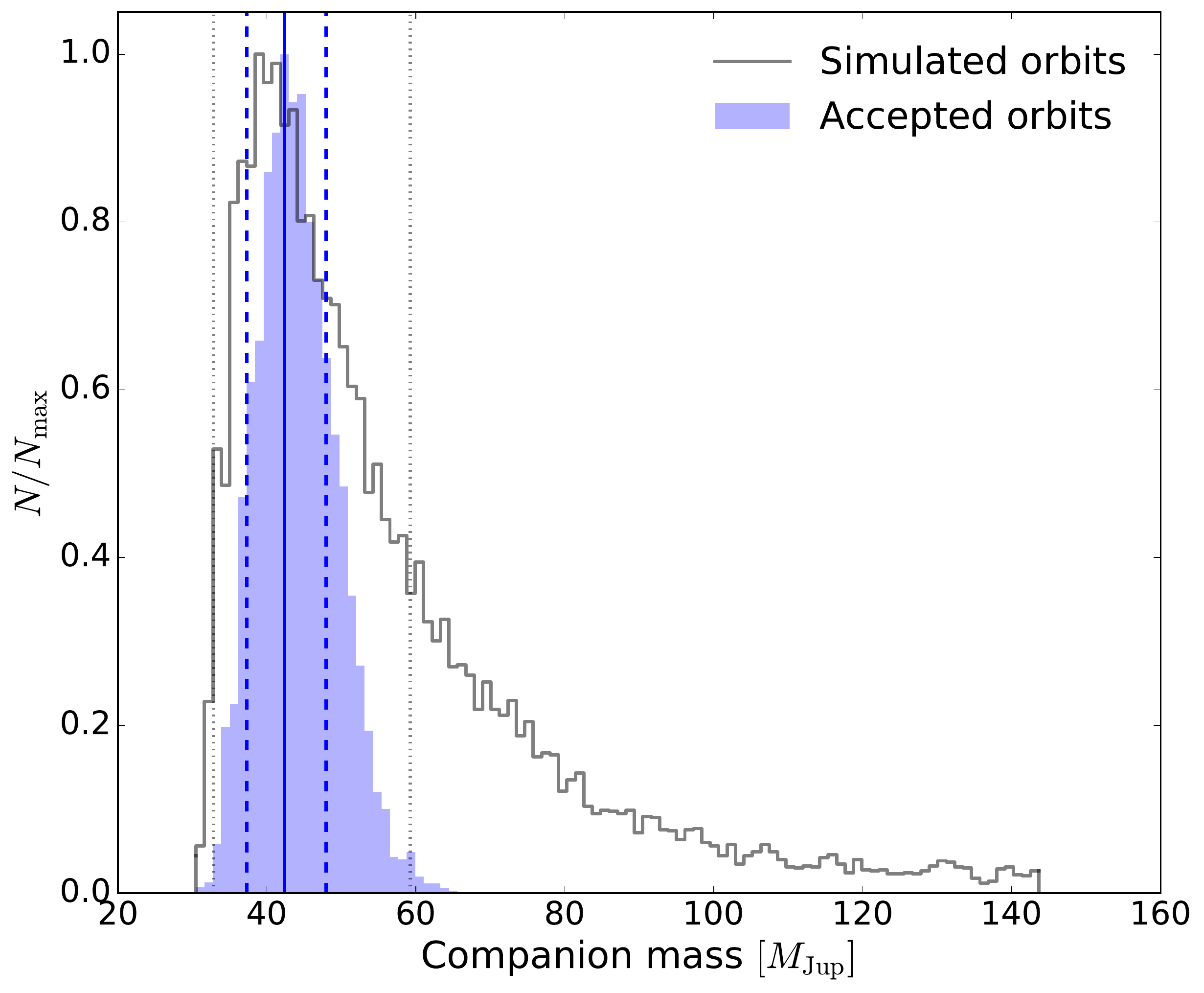}
\caption{Companion mass distribution, where the solid black line shows the $N = 20\,000$ simulated orbits that were best fitted to the relative astrometry from \citet{Bowler+18} and the blue-tinted area only includes the accepted orbits that coincide with the differential proper motion between HIPPARCOS and {\it Gaia}. The blue solid line depicts the median of the most populated bin in the accepted orbits histogram, and the dashed and dotted lines indicate the positions of the $1-\sigma$ uncertainty for the estimated companion mass for the accepted orbits and all orbits respectively, computed as the minimum range that encompasses $68.27 \%$ of the predicted mass-values. }
\label{fig:mass}
\end{figure}

For the allowed orbits, we obtain the respective companion masses, seen in the histogram in Fig.~\ref{fig:mass}. For the most probable value, we find the most commonly populated mass bin and take the median value inside the bin, yielding a companion mass of $M_{\rm c} = 42.4^{+5.6}_{-5.0} \, M_{\rm Jup}$, where the error is taken as the limits of the minimum range that encase $68.27 \%$ of the allowed values. The 20 000 saved orbits in the simulation that were best fitted to the relative astrometry are also shown in Fig.~\ref{fig:mass}, for which we obtain a companion mass of $40^{+19}_{-7} M_{\rm Jup}$, reproducing a similar distribution to the original simulation by \citet{Bowler+18}.

We note that there are differences between the HIPPARCOS proper motions as calculated in \citet{Perryman+97} and \citet{van Leeuwen}, respectively, which exceed their formal error bars. This reflects the difference between the two measured $\Delta \mu$ values plotted in Fig.~\ref{fig:pm}, where the measured differences in proper motions for the \citet{van Leeuwen} solution are measured as $\Delta\mu_{\rm RA} = -1.60 \pm 0.34$ mas yr$^{-1}$ and $\Delta\mu_{\rm DEC} = -1.93 \pm 0.30$ mas yr$^{-1}$. Therefore, while the \citet{Perryman+97} astrometry overlaps with our simulated values as expected, the \citet{van Leeuwen} reduction deviates beyond the error bars in right ascension space. However, in declination space, the two reductions are consistent within error bars, and this is the direction along which most of our expected acceleration takes place over the relevant baseline. The difference between the reductions therefore does not have a significant impact on the mass estimation.  Replacing the proper motion value in declination for the \citet{van Leeuwen} solution yields a companion mass of $39 M_{\rm Jup}$, which is within the error bars of the formally stated result.

We further probe the impact of the updated parallax value from {\it Gaia} DR2 of $\pi_{\rm {\it Gaia}} = 64.06 \pm 0.02$ mas and the original parallax solution for HIPPARCOS by \citet{Perryman+97} of $\pi_{\rm {Perryman}} = 64.54 \pm 0.60$ mas. The change in distance has little impact on the constraints on the dynamical mass, and we subsequently obtain companion masses of $M_{\rm c, \pi_{\it Gaia}} = 42.4^{+5.1}_{-5.1} M_{\rm Jup}$ and $M_{\rm c, \pi_{Perryman}} = 42.5^{+4.2}_{-5.8} M_{\rm Jup}$. We verify from the original catalogues that both of the HIPPARCOS and {\it Gaia} astrometry are calculated using a five-parameter solution, and that the system has been treated as a single star for the reduction \citep[][]{Perryman+97, van Leeuwen, Lindegren+16}. We confirm from the HIPPARCOS Intermediate Data Tool that there were no outliers or other signs of bias to the proper motion in the data. We also check that the correlation between proper motions in right ascension and declination, $\rho(\mu_{\alpha}, \mu_{\delta})$, is consistent between the two catalogues, estimated as $-0.05$ and $-0.03$ for the HIPPARCOS and the {\it Gaia} catalogues,
respectively. 

\section{Discussion \& Conclusions} \label{sec:conc}
\subsection{Previous studies}
\noindent Using the previously calculated orbits for a binary system with long RV trends, we can compare the difference in proper motion, $\Delta \mu$, between the HIPPARCOS and {\it Gaia} epochs in order to further constrain the previous dynamical mass estimates. Here, we employ our method by re-simulating posterior distributions from orbital fits and determine which orbits are consistent with the measured difference in proper motions. \citet{Wielen+99} use a similar approach to identify binaries from long-term ground-based proper motions and HIPPARCOS, arguing that a statistically significant difference in proper motion with respect to the measurement errors could be due to the prevalence of unseen companions. In \citet{Wielen+99}, the proper motion component caused by orbital motion of an intermediate-period companion is assumed to largely cancel out over the very long baseline of the ground-based data, while the HIPPARCOS astrometry retains an orbital motion term in its (approximately instantaneous) proper-motion measurement. The differential proper motion therefore contains a companion signature, although there is considerable ambiguity in the properties of the implied companion. The same procedure could in principle be applied to the {\it Gaia} data relative to ground-based data. However, in the case of GJ 758, we already have prior information on the orbit, so in this case it is better to relate the {\it Gaia} astrometry to the HIPPARCOS astrometry, to get a more precise constraint on the mass without the ambiguities of the original \citet{Wielen+99} approach.

The exploitation of difference in proper motion between HIPPARCOS and {\it Gaia} DR1 was also discussed by \citet{Lindegren+16}. They define the statistic $\Delta Q$ as a measure of the difference in proper motion between the Tycho-{\it Gaia} Astrometric Solution (TGAS) subset of DR1 and the HIPPARCOS catalogue, normalised by the covariances. They divide the sample into two groups; the first one including $\approx 90\, 000$ bona fide single stars with an astrometric five-parameter solution, and the second sample with $\approx 9\, 000$ stars that include solutions accounting for binaries and acceleration. Surprisingly, the distribution in relative frequency over $\Delta Q$ is similar for the two groups, and differs from the expected theoretical distribution. \citet{Lindegren+16} argue that this effect could be an underestimation of the uncertainties in the catalogues, and the fact that most stars are non-single causes the cosmic scatter to be a real effect. Nevertheless, the non-single star sample displays a higher $\Delta Q$ relative frequency, which confirms the sensitivity of the statistic as a probe for duplicity and unseen companions. Indeed, GJ 758 is in the single star sample, displaying a $\Delta Q = 20$, which is on the higher end of the distribution. This is comparable to similar systems with substellar companions that are treated as single stars such as HD 4747, which has a $\Delta Q$-value above 130, and HR 7672 with a value above 300. Therefore, bona fide single stars in the sample with high $\Delta Q$ values require special attention and may be revealed to host substellar companions.

\subsection{Astrometric solution \& primary component}
We note that brightness of GJ 758 with $G = 6.1$ mag is close to the optimal limit for {\it Gaia} \citep[see e.g.][]{Lindegren+18}. For brighter stars, the parameters obtained for the DR2 require careful interpretations. The astrometric excess noise in the DR2 characterises the nature of the solution as a good fit for the source, and is only equals to zero if all observations fit the single-star model well enough. For GJ 758, the quantity is estimated as 0 mas, the same as for other, similar systems such as HD 4747 and HD 4113 that are slightly fainter and thus coincide with the optimal brightness range for {\it Gaia}. In comparison, the slightly brighter system GJ 504 with a magnitude of $G = 5.0$ mag displays an astrometric excessive noise of 0.59 mas. When we apply our proper-motion test to GJ 504 and its companion we are unable to produce trial $\Delta\mu$-values that coincide with the measured values. Although the curvature of the orbit of the companion in the case of GJ 504 is small and the predicted proper motion values are difficult to measure, we cannot exclude the possibility that discrepancy in predicted and measured $\Delta\mu$ stems from the excessive noise in the data for GJ 504 due to its bright nature.  Furthermore, a $\chi^2$ fit to the residuals of the along-scan measurements performed by {\it Gaia} is used as an alternative measure for how well the single-star model fits the source. For GJ 758, the quantity is measured to $\chi^2_{\rm GJ\, 758} \approx 530$, comparable to HD 4747 with $\chi^2_{\rm HD\, 4747} \approx 280$ and significantly lower than the brighter GJ 504 system with $\chi^2_{\rm GJ\, 507} \approx 4600$.

Beyond the solar system, all sources are treated as point objects and single stars in the {\it Gaia} DR2, and their motions are described by a five-parameter model \citep{Lindegren+18}. Therefore, it is the photocentre that is used for the astrometric solution and not the centre of mass.  We scrutinise the {\it Gaia} source catalogue and find no available photometric variability flags that may bias the astrometric solution. Nevertheless, the orbital motion may affect the astrometric solution, albeit the effect is expected to be small given the low mass ratio of $q = 0.04$ and the relative motion of the companion during the {\it Gaia} observations. 

We assumed a primary stellar mass of $0.97 \pm 0.02 \, M_\odot$ in order to be consistent with the simulations by \citet{Bowler+18}. The value for the stellar mass stems from \citet{Vigan+16}, where evolutionary models by \citet{Bressan+12} were employed with an assumed age of $2.2 \pm 1.4$ Gyr. However, the dynamical mass for the companion implies an older age of $\gtrsim 8$ Gyr, which in turn would be more consistent with a smaller stellar mass according to the models, bringing the components closer to each other in the mass-luminosity diagram. We therefore probe the effect of having a lower stellar mass of $\approx 0.91\,M_\odot$ , to be more consistent with an older age for the primary, and increase the uncertainty to $\pm 0.1\,M_\odot$. We note that this has little impact on our results, with a slightly lower companion mass of $41.5^{+5.6}_{-4.5}\,M_{\rm Jup} $, which is well within the error bars of the formally stated companion mass.

\subsection{Conclusions and outlook}
\begin{figure}
\includegraphics[width=\columnwidth]{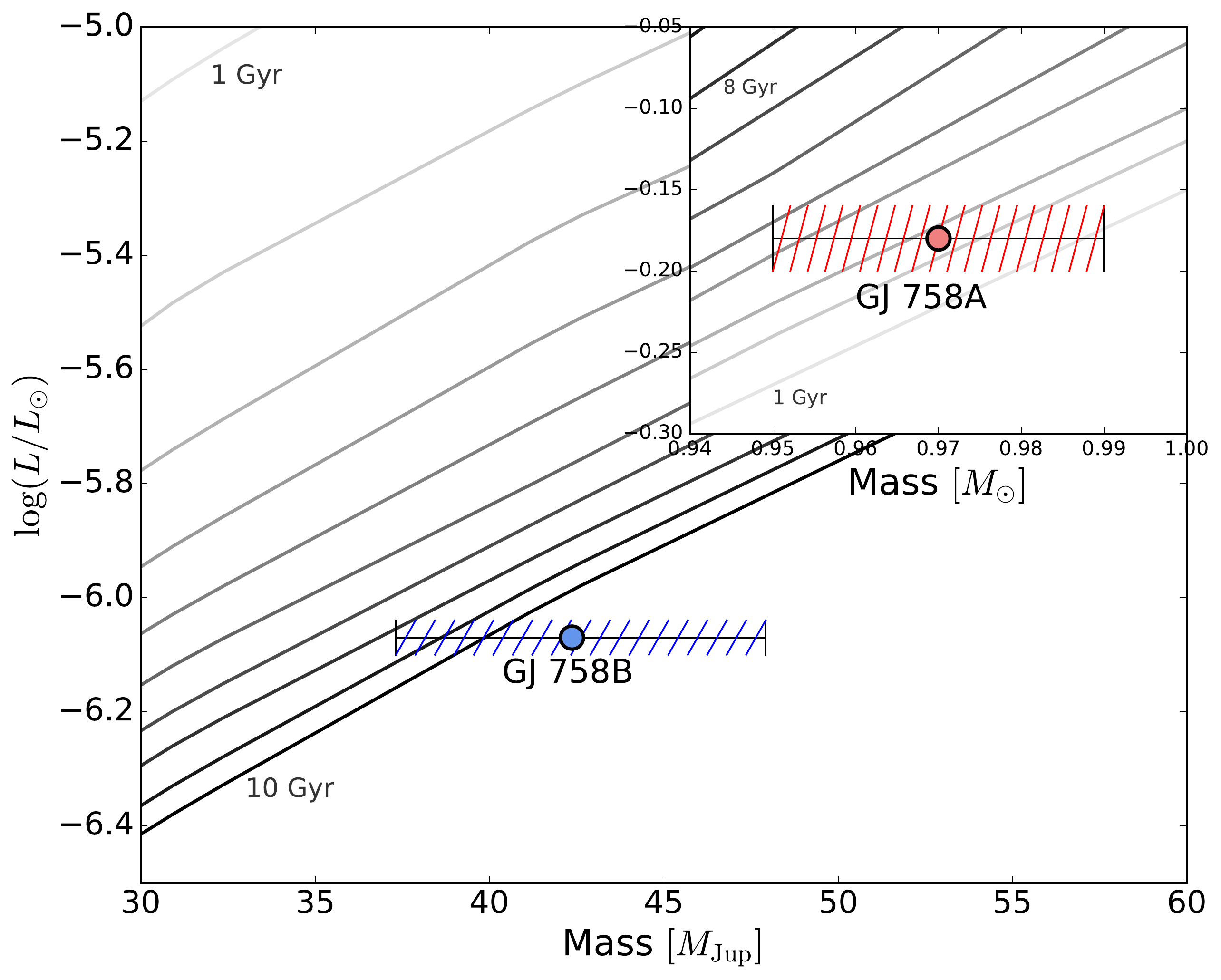}
\caption{Mass-luminosity diagram for GJ 758B with COND \citep{COND} isochrones spanning 1 to 10 Gyr, going from lighter to darker shaded lines for the youngest isochrone to the oldest. The dynamical mass is best fitted to an old age, $\geq$ 8 Gyr, consistent with the current age estimations of the host star, yet displaying a discrepancy in isochronal age.}
\label{fig:ML}
\end{figure}

We obtain a new, more precise dynamical mass-estimate for GJ 758B of $42.4^{+5.6}_{-5.0} \, M_{\rm Jup}$, which can be used to test substellar evolutionary models. We compare GJ 758B in a mass-luminosity diagram using isochrones ranging ages 1 - 10 Gyr from the COND evolutionary models \citep{COND} and the bolometric luminosity of $\log(L/L_\odot) = -6.07 \pm 0.03$ \citep{Bowler+18}, shown in Fig.~\ref{fig:ML}. The models predict an age of $\geq 8 $ Gyr for the companion, which is older than previous estimates for the primary host. This age discrepancy and uncertainty is similar to that of the GJ 504 system, for example, which also hosts a companion with a long-standing ambiguity in its mass \citep[e.g.][]{dOrazi+17}, thus further illustrating the difficulty of determining ages of intermediate-age Sun-like stars, and the importance of dynamical mass measurements for their companions. GJ 758B thereby joins a growing list of cool, T-type substellar companions on medium long periods of $\approx 30 - 100$ years, imaged around solar-type main sequence stars such as GJ 504b, HD 4113C, and HD 4747B \citep[][]{Kuzuhara+13, Cheetham+17, Crepp+18}.

Due to the favourable orbit and the observational data available for GJ 758B, the method we employ works well. For the similar systems described above, only HD 4747B has such an orbit for which $\Delta \mu$ can be properly estimated. However, the period of HD 4747B is $\approx 38$ years, and already has a well-constrained mass of $65.3^{+4.4}_{-3.3} \, M_{\rm Jup}$. This means that we can apply a sanity check to our methodology by applying the same method as described above to HD 4747B, in order to check if the precise $\Delta \mu$ prediction allowed by the existing orbital fit matches the {\it Gaia} DR2 data. We confirm that this is indeed the case, with a measured $\Delta \mu_*$ of 1.63 $\pm$ 1.10 mas yr$^{-1}$ in right ascension and 4.33 $\pm$ 0.81 mas yr$^{-1}$ in declination for HD 4747. The predicted $\Delta\mu_*$-values for HD 4747 range from 1 to 6 mas yr$^{-1}$ with a median of 2.7 mas yr$^{-1}$ in right ascension  and from 2 to 6 mas yr$^{-1}$ with a median of 3.38 mas yr$^{-1}$ in declination. The trial orbits consistent with the measured $\Delta\mu_*$-values yield a mass of $M_{\rm HD\,4747B} = 64.1^{+4.6}_{-2.4}\, M_{\rm Jup}$. For the case of HD 4747, the astrometry is consistent between the two reductions, and both are consistent with our simulated predictions. We suggest that future studies involving the $\Delta \mu$ between HIPPARCOS and {\it Gaia} should consider both existing HIPPARCOS reductions, note whether there are systematic differences between them, and if so, compare the resulting $\Delta \mu$ values to see if the specific reduction has an effect on the result.


A reasonable assumption is that both components in the GJ 758 system are coeval, and that the age of the companion can be used to place constraints on the host and thereby entire the system. However, at these very low substellar masses, evolutionary models are not well-constrained and generally underpredict masses based solely on the brightness of an object. Therefore, precise dynamical mass-estimates of substellar objects such as GJ 758B can be used as touchstone objects for calibrating evolutionary models at substellar masses. The fact that GJ 758B is likely to be one of the lowest-mass and coldest substellar companions that has both a long RV baseline and has been resolved in direct imaging makes it an important benchmark object for calibrating models down towards the planetary-mass regime. The observations demonstrate the power of {\it Gaia} for constraining the properties of wide companions, particularly with the long baseline afforded when combining its astrometry with that of HIPPARCOS, leading to substantial curvature in the orbit and thus a large $\Delta \mu$. Continued observations with \emph{Gaia}, further high-contrast imaging, and RV monitoring are also likely for even better constrained orbit and dynamical mass.

The presented results provide indications for promising future directions in exoplanet characterisation. The atmosphere and brightness of GJ 758B are broadly equivalent to giant planets at younger ages of a few tens or hundreds of millions of years \citep{COND}. Such planets are already potentially detectable with direct imaging, and with the increased time baseline and better precision of the eventual full {\it Gaia} release, sensitivity to such masses will also be readily provided for large numbers of targets. Therefore, our ability to simultaneously constrain the atmospheric and physical properties of wide-planet and brown dwarf companions will substantially increase in the coming years through the combination of high-contrast imaging and astrometric data as combined in this study.

\begin{acknowledgements} 
We thank the referee, Johannes Sahlmann, for the very useful comments which helped to improve the manuscript. We also thank Micka\"{e}l Bonnefoy and Laetitia Rodet for useful discussion. This study made use of the CDS, NASA-ADS and \emph{Gaia} archive online resources, and of data from the European Space Agency (ESA) mission {\it Gaia} (\url{https://www.cosmos.esa.int/gaia}), processed by the {\it Gaia} Data Processing and Analysis Consortium (DPAC, \url{https://www.cosmos.esa.int/web/gaia/dpac/consortium}). Funding for the DPAC has been provided by national institutions, in particular the institutions participating in the {\it Gaia} Multilateral Agreement. M.J. gratefully acknowledges funding from the Knut and Alice Wallenberg foundation.

\end{acknowledgements}

\bibliographystyle{aa-note} 
\bibliography{aa33309bib}      

\end{document}